\documentclass{ieeeaccess}
\usepackage{cite}
\usepackage{amsmath,amssymb,amsfonts}
\usepackage{algorithmic}
\usepackage{graphicx}
\usepackage{textcomp}
\def\BibTeX{{\rm B\kern-.05em{\sc i\kern-.025em b}\kern-.08em
    T\kern-.1667em\lower.7ex\hbox{E}\kern-.125emX}}
\begin{document}
\history{Date of publication xxxx 00, 0000, date of current version xxxx 00, 0000.}
\doi{10.1109/ACCESS.2017.DOI}

\title{DCAN: Diversified News Recommendation with Coverage-Attentive Networks}
\author{\uppercase{Hao Shi}\authorrefmark{1},
\uppercase{Zi-jiao Wang\authorrefmark{2}, and Lan-ru Zhai}\authorrefmark{1}}
\address[1]{
Department of Electronic Information, Jinzhong Vocational and Technical College, Jinzhong, CO 030600, China}
\address[2]{Taiyuan Tax Service, State Administration of Taxation, Taiyuan 030000, China)}

\markboth
{HAO SHI \headeretal: Preparation of Papers for IEEE TRANSACTIONS and JOURNALS}
{HAO SHI \headeretal: Preparation of Papers for IEEE TRANSACTIONS and JOURNALS}

\corresp{Corresponding author: Zijiao Wang (e-mail: 540360789@qq.com).}

\begin{abstract}
Self-attention based models are widely used in news recommendation tasks. However, previous Attention architecture does not constrain repeated information in the user's historical behavior, which limits the power of hidden representation and leads to some problems such as information redundancy and filter bubbles. To solve this problem, we propose a personalized news recommendation model called DCAN.It captures multi-grained user-news matching signals through news encoders and user encoders. We keep updating a coverage vector to track the history of news attention and augment the vector in 4 types of ways. Then we fed the augmented Coverage vector into the Multi-headed Self-attention model to help adjust the future attention and added the Coverage regulation to the loss function(CRL), which enabled the recommendation system to consider more about differentiated information. Extensive experiments on Microsoft News Recommendation Dataset (MIND) show that our model significantly improve the diversity of news recommendations with minimal sacrifice in accuracy.
\end{abstract}


\titlepgskip=-15pt

\maketitle

\section{Introduction}
\label{sec:introduction}
\PARstart{T}{he}News Recommender System (NRS) has been widely used in online news websites \cite{b14},\cite{b31}. It helps users to find news of interest in huge amounts of data. In most studies, accuracy is the primary or even the only goal of news recommendation. However, improving the diversity of recommended content is essential for enhancing \cite{b1},\cite{b48},\cite{b37} user experience. The excessive pursuit of accuracy indirectly leads to information redundancy in recommended content \cite{b39}, users' boredom \cite{b23},\cite{b25} and filter bubbles \cite{b17}, \cite{b35},\cite{b46}, which have many negative effects on society. 

To address these issues, some studies \cite{b1},\cite{b48}have used diversity to improve user engagement and user satisfaction in recommendation scenarios. However, many existing diversity-oriented news recommendation methods \cite{b21},\cite{b41},\cite{b45} perform poorly, and only a few methods \cite{b36},\cite{b43} have achieved good results. Moreover, most of these algorithms ignore the use of the Coverage mechanism to improve recommendation diversity. The Coverage mechanism is common to machine translation and has achieved satisfactory results. Since the Coverage vector contains historical information, it can be used to model the information that is different from the historical behaviour in candidate news and improve the diversity of recommendation. 

In this paper, we propose Deep Coverage Attentive Networks(DCAN), which integrates the Coverage attributes into recommendations in an end-to-end and extensible way. Specifically, we model user behaviour as a Coverage sequence in the latent space, and then augment the Coverage sequence to mine multi-view Coverage information. Afterwards, multiple augmented Coverage sequences are embedded into different self-attention headers of the model. Finally, we also configure the corresponding regularizer for the target function according to the different Coverage sequences. Experiments show that our work can effectively improve the diversity of news recommendations with minimal loss of accuracy. Experiments show that our work can effectively improve the diversity of news recommendations with minimal sacrifice in accuracy. Our contributions are as follows.

• We propose a novel DCAN model, which is a news recommendation model combining diversification and relevance. 

• Coverage mechanism in machine translation is used to improve multiple self-attention mechanisms and objective functions

• Experiments on news datasets demonstrate the effectiveness of our model. We also investigate the impact of various novel modules in the model through ablation experiments

\section{RELATED WORK}
\subsection{News recommendation}
The core problem of news recommendation is learning the representation of news and users and ranking candidate news based on their representation \cite{b41}. Modeling news representations are generally from the title, category, entity, keyword, abstract, tag, etc, while the information of modeling users includes news click, user ID, Time, Location, Non-click, Finish, Quick Close, Share, Dislike, etc.\cite{b55}. In recent years, several news recommendation methods based on deep learning techniques have been proposed and achieved better performance than traditional methods. DKN \cite{b31} NPA\cite{b18} and \cite{b45},\cite{b53},\cite{b51} use CNN and personalized attention for representation learning, NRMS \cite{b34} use self-attentional networks for news representation, Plm-nr \cite{b15} UNBERT \cite{b52} uses the pre-trained BERT model for feature representation. Slightly different from these methods, our model learns news representations through transformers and forms user representations using multi-head self-attention with Coverage embedding. 
\subsection{Diversity recommendation}
Diversity recommendation is often regarded as a bi-criteria optimization problem, which seeks to maximize the relevance of the recommended list while minimizing the redundancy among the recommended items \cite{b5}. Diversity recommendations usually have two solutions: aggregate-level diversity \cite{b56}and individual diversity. The former aims to increase the differentiation of all recommended content and exposure to long-tail content \cite{b4},\cite{b16},\cite{b9}, while the latter aims to improve the intra-user diversity \cite{b22},\cite{b24}. Our work focuses on the latter. 

The diversity recommendation methods can be divided into two categories:offline optimization and online optimization. Offline non-interactive methods usually generate recommendations based on user historical interactions and then re-rank them. These include post-processing methods based on heuristics \cite{b42}and determinantal Point Process \cite{b16} \cite{b29} \cite{b47}, and Learning to Rank (LTR) \cite{b28} \cite{b27} generation ranking methods. Online approaches are to update recommendation policies based on real-time interactions between users and recommendation systems, such as contextual bandits \cite{b9},\cite{b10}and deep reinforcement learning \cite{b57}. 

\subsection{Coverage mechanism}
Coverage mechanisms are generally used to mitigate over-translation and under-translation problems in neural machine translation. \cite{b32} summed the attention distribution during decoding to obtain a Coverage vector. This vector was used to track the attention history and to compute a new attention distribution at the next time step. \cite{b19} proposed a Coverage-based loss function to penalize attentional repetition in text summarization during optimization. 
Coverage mechanisms play a role in enhancing diversity in recommender systems. \cite{b11} proposed a Coverage method based on a probability model for diversity reordering algorithm.\cite{b2} represents news as nodes in a similarity graph, and improves recommendation diversity by recommending subsets of news that are positively rated by users and have low Coverage values. 

\section{Our Approach}
In this section, we first formulate the problem of news recommendation, then detail our DCAN including the News Encoder, User Encoder and the prediction module(Figure 1), and finally elaborate on a model training method. 
\subsection{Problem Statements}
For a given user u and candidate news v, we denote the history of news clicks for user u as $V^{u}=\left[v_{1}^{u}, \ldots v_{\left|V^{u}\right|}^{u}\right]$. Our goal is to predict the next news that user u is likely to interact with based on the given history $V_u$ and to maximize the diversity of recommended content. The output of DCAN is a matching score, which represents the probability that user u will click on the candidate news.
\subsection{News Encoder}
The news encoder is used to learn news representations from features such as news title, entity, abstract, category, etc. This article only uses the feature of the news title, which consists of a series of words. As shown in Figure 2, for any news $\mathrm{v}_{\mathrm{i}}^{\mathrm{u}}=\left[\mathrm{w}_{\mathrm{i} 1}^{\mathrm{u}}, \ldots \mathrm{w}_{\mathrm{ij}}^{\mathrm{u}}, \ldots \mathrm{W}_{\mathrm{i}\left|\mathrm{v}_{\mathrm{i}}^{\mathrm{u}}\right|}\right]$, $w_{ij}^u$ represents the jth word. Inspired by the news encoder \cite{b34}, we first use a word embedding layer to convert word sequences in news titles into the embedding vector sequence $\mathrm{E}^{\mathrm{u}}=\left[\mathrm{e}_{1}^{\mathrm{u}}, \mathrm{e}_{2}^{\mathrm{u}}, \ldots \mathrm{e}_{\left.\mid \mathrm{E}^{\mathrm{u}}\right]}^{\mathrm{u}}\right]$. Then we use Transformer \cite{b6}encoder to capture the context of words and build a d-dimensional vector to represent the news. We denote the sequence of news clicks learned by the news encoder as $\mathrm{R}^{\mathrm{u}}=\left[\mathrm{r}_{1}^{\mathrm{u}} \mathrm{r}_{2}^{\mathrm{u}}, \ldots, \mathrm{r}_{\left.\mid \mathrm{R}^{\mathrm{u}}\right]}^{\mathrm{u}}\right]$, and the candidate news representations are denoted as $r_c$. 

\subsection{User Encoder}
The User Encoder module is used to learn the user's representation from the news that the user browses. It consists of five parts.
\subsubsection{User Encoder Input Layer}
{After $V_u$ has been converted into $R_u$ by news encoder, we feed $R_u$ into the model. As in previous work \cite{b8},\cite{b30}, our User Encoder is a fixed-length model with length N. Therefore, $R_u$ needs to be converted to $\widetilde{\mathrm{R}^{\mathrm{u}}}=\left[\mathrm{r}_{\left|\mathrm{R}^{\mathrm{u}}\right|-\mathrm{N}+1}^{\mathrm{u}}, \ldots, \mathrm{r}_{\left|\mathrm{R}^{\mathrm{u}}\right|-1}^{\mathrm{u}}, \mathrm{r}_{\left|\mathrm{R}^{\mathrm{u}}\right|}^{\mathrm{u}}[\operatorname{mask}]\right]\in \mathbb{R}^{N\times d}$. If the history is less than N-1, it will be padded with a special token [PAD].}
\subsubsection{Coverage Feature representation module}
In this part, we constructed the Coverage feature representation sequence $\operatorname{Cov}^{\mathrm{u}}=\left[\mathrm{C}_{1}^{\mathrm{u}}, \mathrm{C}_{2}^{\mathrm{u}}, \ldots \mathrm{C}_{\left|\operatorname{Cov}^{\mathrm{u}}\right|}^{\mathrm{u}}\right]$ based on $R_u$, and satisfied $\mathrm{c}_{1}^{\mathrm{u}}=\mathrm{r}_{1}^{\mathrm{u}}$, $\mathrm{c}_{\mathrm{i}}^{\mathrm{u}}=\mathrm{c}_{\mathrm{i}-1}^{\mathrm{u}}+\mathrm{r}_{\mathrm{i}}^{\mathrm{u}}$ and $c_{i} \in \operatorname{Cov}_{i}$. For a fixed-length model of length N, the Coverage sequence is $\widetilde{\operatorname{Cov}^{\mathrm{u}}}=\left[\mathrm{C}_{\left|\operatorname{Cov}_{\mathrm{u}}\right|-\mathrm{N}+1}^{\mathrm{u}}, \ldots, \mathrm{c}_{\left|\operatorname{Cov}^{\mathrm{u}}\right|-1}^{\mathrm{u}}, \mathrm{c}_{\left|\operatorname{Cov}^{\mathrm{u}}\right|}^{\mathrm{u}}[\text { mask }]\right]\in \mathbb{R}^{\mathrm{N} \times \mathrm{d}}$. 
\subsubsection{Coverage Augmentation Module}
To obtain different views of the Coverage properties, we take four types of data augmentation: Decay, Circle, Log, Gamma. The first one is the Decay encoder ($C^{Decay}$), which converts $c_i^u$ to $c_i^{Decay}$ using the following equation. \begin{equation}
    c_{i}^{\text {Decay }}=r_{i}+\eta r_{i-1}+\eta^{2} r_{i-2}+\ldots+\eta^{i-1} r_{1}, \eta \in[0,1]
\end{equation}
The Circle encoder ($C^{Circle}$) converts the $c_i$ into hidden vectors $c_{i}^{\text {circle }}=\left\{c_{i}^{\sin }, c_{i}^{\cos }\right\} \in \mathbb{R}^{d}$ by the following formula.
\begin{equation}
    c_{i, 2 a}^{\sin }=\sin \left(\frac{c_{i, 2 a} \times i}{\text { freq }^{\frac{2 a}{d}}}\right), c_{i, 2 a+1}^{\cos }=\sin \left(\frac{c_{i, 2 a+1} \times i}{\text { freq }^{\frac{2 a+1}{d}}}\right)
\end{equation}
where is $c_{i, a}$ the ath value of the $c_i$ vector and freq is an adjustable parameter. Similarly, the Log encoder ($C^{Log}$) and Gamma encoder ($C^{Gamma}$) convert $c_i$ to $\mathrm{c}_{\mathrm{i}}^{\log } \text { and } \mathrm{c}_{\mathrm{i}}^{\mathrm{Gamma}}$ by the following equations, respectively.
\begin{equation}
    c_{i, a}^{\log }=\log \left(1+\frac{c_{i, a}}{\text { freq }^{\frac{a}{\mathrm{~h}}}}\right), \quad c_{i, a}^{G a m m a}=\beta c_{i, a} * e^{\left(-\beta c_{i, a}\right)}
\end{equation}
Where $\beta$ is the scale hyperparameter. After augmentation, the , each augmentation provides a unique view of the Coverage data. For example, Circle captures periodic features. Decay's decay factor makes it more focused on recent news. The Log can alleviate the problem that large values have too much influence on the results. Finally, the Gamma encoder keeps the best Coverage value in a moderate range. 
\subsubsection{Coverage embedded Self-Attention Structure}
There are several options for embedding encoding functions into attention headers. Self-attention based Models \cite{b7},\cite{b44}, \cite{b8}take the sum of $\widetilde{\mathrm{R}_{\mathrm{u}}}$ and position encoding $\widetilde{\operatorname{Cov}^{\mathrm{u}}}$ as the input of [Query][Key]and[Value]. Advanced transformers \cite{b50} position encoding is replaced by features and only [key] is embedded, or the feature of representations are embedded in [query] and [key]\cite{b30}. 
In our work, we first divided the vectors $\mathrm{C}_{\mathrm{i}}^{\text {Circle }} \mathrm{C}_{\mathrm{i}}^{\text {Decay }} \mathrm{C}_{\mathrm{i}}^{\text {Log }} \mathrm{C}_{\mathrm{i}}^{\text {Gamma }}$ by the historical click times i to calculate the average value, then inserted the obtained vectors into [Value]  as shown in Figure 3. Since Coverage can track attention history, the [Value] used for decoding learns information that is different from historical behaviour. 
\subsubsection{Stacking Self-Attention Blocks}
The part after self-attention is similar to \cite{b8},\cite{b30}. We feed the self-attention output into a feed-forward network (FFN), and then stack the self-attention and FFN L times. Like \cite{b8},\cite{b44}, we build a residual connection for each stacking module and then use layer normalization \cite{b26} to facilitate training.\begin{equation}
    \begin{array}{c}
\text { FFN }(\mathrm{x})=\mathrm{GELU}\left(\mathrm{xM}^{1}+\mathrm{b}^{1}\right) \mathrm{M}^{2}+\mathrm{b}^{2} \\
\mathrm{y}=\operatorname{layerNorm}(\mathrm{x}+\operatorname{Dropout}(\operatorname{Attention}(\mathrm{x}))) \\
\mathrm{z}=\operatorname{layerNorm}(\mathrm{y}+\operatorname{Dropout}(\operatorname{FFN}(\mathrm{y})))
\end{array}
\end{equation}
Where GELU is the Gaussian Error Linear Unit and $x \in \mathbb{R}^{1 \times d}$ is the output of the Coverage embedded multi-head Attention(CMA). $M^{1} \in \mathbb{R}^{d \times 4 d}$, $b^{1} \in \mathbb{R}^{4 d}$ $\mathrm{M}^{2} \in \mathrm{D}^{\mathrm{d} \times 4 \mathrm{~d}}$ and $b^2\in \operatorname{\mathbb{R}}^{\mathrm{d} \times 4 \mathrm{~d}}$ is a learnable parameter. 
\subsection{Prediction Module}
Given the output $\mathrm{O}=\left[\mathrm{o}_{1}, \ldots, \mathrm{o}_{\mathrm{i}}, \ldots, \mathrm{o}_{\mathrm{N}}\right] \in \mathbb{R}^{\mathrm{N} \times \mathrm{d}}$ of the last layer , we obtain the news score distribution for each position by using the following equation.
\begin{equation}
P(V \mid r, c)=\operatorname{softmax}\left(G E L U\left(o_{i} M^{p}+b^{p}\right) M^{T}+b^{o}\right)
\end{equation}
Where $M \in \mathbb{R}^{d \times d}, b^{p} \in \mathbb{R}^{d} \text { and } b^{0} \in \mathbb{R}^{|R|}$ are the learnable parameters, and $M^{T} \in \mathbb{R}^{\mathrm{d} \times\left|\mathrm{r}_{\mathrm{c}}\right|}$ is the candidate news embedding. $\mathrm{P}\left(\mathrm{r}_{\mathrm{i}}=\mathrm{r}_{\mathrm{i}} \mid \mathrm{r}, \mathrm{c}\right)$ is the probability that an news at position i is news r.
\subsection{Model Training }
Overall, to provide recommendations with both accuracy and diversity, we calculate the following mixed loss functions to train the DCAN model.
\begin{equation}
\mathrm{L}=\mathrm{L}_{\text {main }}+\gamma \mathrm{L}_{\text {diverse }}
\end{equation}
where $L_{mian}$ is the click prediction loss function, and $L_{diverse}$ is a Coverage-based diversification-oriented regularization loss (DOR).
\subsubsection{Main loss}
We train the model using the techniques in \cite{b8},\cite{b19} .We feed $V_u$ into news enconder to generate $R_u$ and $C_u$.We sample the news subsequence $\mathrm{R}^{\mathrm{u}, \mathrm{s}} \in \mathrm{R}_{\mathrm{u}}$ and $C^{u, s} \in C_{u}$ respectively and convert $\mathrm{R}^{\mathrm{u}, \mathrm{s}}$ to $\overline{\mathrm{R}}^{\mathrm{u}, \mathrm{s}}$ by randomly masking $\overline{\mathrm{R}}^{\mathrm{u}, \mathrm{s}}$ with probability $\rho$. Finally, we feed $\overline{\mathrm{R}}^{\mathrm{u}, \mathrm{s}}$ and $C_{u}$  to our model and get the probability that news at position i is $\overline{\mathrm{R}}_{\mathrm{i}}^{\mathrm{u}, \mathrm{s}}$. We calculate the loss as follows:
\begin{equation}
    L_{\text {main }}=-\sum_{\overline{\mathrm{R}}_{\mathrm{i}}^{\mathrm{u}, \mathrm{s}}\text { is masked }}  \log P\left(\overline{\mathrm{R}}_{\mathrm{i}}^{\mathrm{u}, \mathrm{s}}=\overline{\mathrm{R}}_{\mathrm{i}}^{\mathrm{u}, \mathrm{s}} \mid \overline{\mathrm{R}}^{\mathrm{u}, \mathrm{s}}, \mathrm{C}^{\mathrm{u}, \mathrm{s}}\right)
\end{equation}
\subsubsection{Diverse-Oriented Regularization}
Finding differentiated recommendation information is an effective way to diversify recommended content, but in fact, it is difficult to estimate the feature distribution of diverse information. To solve this problem, the training objective is to enlarge the distance between the output feature distribution and Coverage feature distribution. We add a regular function based on the Coverage mechanism to the original loss function. Specifically, the Coverage vector $c_i$ is augmented to $C_i^{Decay} ,C_i^{Circle}, C_i^{Log}, C_i^{Gamma}$ according to the Coverage embedded multi-head Attention,the augmented vectors are summed to $C_i^{\mathrm{COV}} $.If the corresponding coverage vector has been embedded with multiple attention 
$\Phi$=1 otherwise $\Phi$=0.
\begin{equation}
    \mathrm{C}_{\mathrm{i}}^{\text {Cov }}=\mathrm{C}_{\mathrm{i}}+\Phi \mathrm{C}_{\mathrm{i}}^{\text {Decay }}+\Phi \mathrm{C}_{\mathrm{i}}^{\text {Circle }}+\Phi \mathrm{C}_{\mathrm{i}}^{\mathrm{Log}}+\Phi \mathrm{C}_{\mathrm{i}}^{\text {Gamma }}
\end{equation}

We calculated the L2 norm \cite{b26} of the output vector $o_i$ and Coverage vector $C_i^{\mathrm{COV}}.$
\begin{equation}
    \bar{o}_{i}=\frac{o_{i}}{\left\|o_{i}\right\|_{2}},  
    \overline{C_{i}^{C O V}}=\frac{c_{i}^{C O V}}{\left\|c_{i}^{C O V}\right\|_{2}}
\end{equation}
Then we define the diverse loss with the help of  Mean Squared Error:
\begin{equation}
    \mathrm{L}_{\text {diverse }}=-\left\|\overline{o_{i}}-\overline{c_{i}^{C O V}}\right\|_{2}^{2}
\end{equation}

\section{EXPERIMENTS}
\subsection{Datasets}
We evaluated our model on the real-world news recommendation dataset MIND \cite{b31}, which consists of anonymous behaviour logs from Microsoft News. There are two versions of the MIND datasets named MIND-large and MIND-small. Where MIND-small is a small version of MIND-large by randomly sampling daily behaviour with equal probability. The basic statistics for these two datasets are shown in Table 1. For these two data sets, we generate training samples from click history and impression logs according to the format given in the MIND paper\cite{b31}. Impression logs record the news articles that readers visit or click on at a specific time \cite{b31}. We grouped the interactions according to user ids and then sorted them according to timestamps, forming a sequence for each user.
\subsection{Assessment}
To evaluate the performance of the recommendation model, we keep the last news of each user's click sequence as test data, which is widely used in \cite{b33},\cite{b40},\cite{b44}. The item just before the last item is considered to be validation data. We train with the rest of the news. We use the rest of the news as a training set. Inspired by \cite{b8},\cite{b30},\cite{b44},\cite{b49}, the model takes 100 randomly sampled news that users have not yet interacted with as negative samples and pairs them with ground truth. Negative sampling correlates with popularity.To score the ranking list, we use three metrics: Accuracy (AUC), Normalized Discounted Cumulative Gain (NDCG), and Diversity Score(DIV). We set diversity as the difference of news pairs under a certain cutoff (k) in a user's recommendation list. Thus, we use the intra-column similarity (ILS) \cite{b22},\cite{b24}as the diversity metric. AUC, NDCG@K, and DIV@K are all designed to have larger values when ground truth news ranks higher in the top-K list. We compare our models with baselines in accuracy and diversity. We expect to increase model diversity while maintaining as much accuracy as possible. 
\subsection{Baseline model and implementation details}
To verify the effectiveness of our method, we compared it with the following representative baselines: MMR \cite{b24} DPP \cite{b18},\cite{b29},\cite{b47}, BERT4Rec \cite{b8}, NRMS \cite{b34} and Fastformer \cite{b58}. In our experiments, we use Glove \cite{b20} word embedding for initialization. We implemented DCAN with PyTorch. In our experiments, we use the mind-small dataset to determine parameter settings, and then we train and evaluate on both small and large datasets. For our own model, the word embedding dimension size is 200, and news embedding dimension size is d = 128, the number of self-attention heads is n = 8, the dropout rate is 0. 2, and batchsize is 64. For our model, we set the word embedding dimension size $d_{word}$ = 200, and news embedding dimension size d = 128, all the number of self-attentive network heads n = 8, the dropout rate 0. 2, and the batchsize 64. The maximum news sequence length is N = 50 and the number of stacked layers is L = 2. We embed four augmented Coverage vectors into the model, and the rest of the heads do not add any Coverage Embedding. Whenever a Coverage vector is embedded , we add the corresponding loss regulation to the $L_{Diversity}$. We use Adam \cite{b3} for model optimization. To ensure fairness, we tune all models under the same set of hyperparameters. Each experiment was repeated 5 times. 
\subsection{Results}
Table 2 shows the results for all models on two datasets (mind-small and mind-large). It also shows the comparison between our model and the strongest baseline. Overall, our model achieves the best performance on all diversity metrics, while the click prediction accuracy (AUC) and recommendation accuracy (NDCG) losses are within a reasonable range(-1. 55$\%$ and -3. 38$\%$ on average). For the evaluation of DIV@10, DIV@20 and DIV50, our model achieved 4. 36$\%$, 3. 7$\%$ and 2. 5$\%$ improvement over the strongest baseline, respectively. We observe that the overall performance of the attention-based methods is better than the traditional recommenders. Since most of our baseline models utilize the multi-head attention mechanism,We also find that the performance improvement comes from our proposed novel Coverage embedded multi-head attention and Diverse-Oriented Regularization loss function.
 
 In order to analyze the impact of different Coverage embedding and regularization functions in our model, we conduct an ablation study on the MIND-small dataset. We unified all the other hyperparameters. Table 3 
shows the performance of our default DCAN method and its variants. DCAN consists of 4 kinds of coverage vectors include $C^{Decay} ,C^{Circle}, C^{Log}, C^{Gamma}$ and  4 kinds of DORs includes $DOR^{Decay},DOR^{Circle},
DOR^{Log},DOR^{Gamma}$.Our model variants are named DCAN. When we remove some side information, we use a 
minus sign in front of the coverage vectors and DORs, e.g.$-C^{Decay}-DOR^{Decay}$ is DCAN without $-C^{Decay}$ and $-DOR^{Decay}$.The results show that the importance of each kind of embedding seems to be different. In general, the performance of multiple embedding is generally greater than that of individual embedding. The Circle embedding seems to have a greater impact on the DIV than other kinds of embedding. This suggests that although users have relatively fixed reading preferences for news, there are inherent differences in short-term news browsing (within one day), so the modeling effect of short-term Coverage embedding such as Decay and Gamma is not significant. This suggests that diversity modeling needs to take into account the reading habits of users and flexibly select different Coverage embedding.

In addition, the further analysis of self-attention showed that with the increase of the number of heads, $h=\{8,10,20,25\}$, the diversity fluctuated continuously with no obvious trend. It can be seen that the types of embedding vector and loss function play an important role in the training process, while the variation of attention heads does not significantly improve diversity performance. 
\section{CONCLUSION}
In this paper, we propose a novel method named Deep Coverage Attentive Networks for diversified news recommendations. The core of our method is to provide unique position embedding for the self-attention module by using multiple augmentation methods of Coverage information in the user encoder and adding a corresponding Coverage regularizer into the objective function at the same time. Experiments on real datasets show that our model outperforms state-of-the-art baselines in diversity recommendation. Through extensive ablation studies, we also found that the Coverage augmentation embedding scheme needs to fully consider user preferences. 
In the future, we will improve our method in the following directions. First, we'll try embedding the Coverage attribute into other Transformer variants to see what happens. Second, we consider introducing more features (such as news types, news sentiment, etc. )and techniques (such as contrast learning) to improve diversity.  We will study the diversity feature representation method. Finally, we will further study the diversity representation method(such as GNN).

\begin{IEEEbiography}[{\includegraphics[width=1in,height=1.25in,clip,keepaspectratio]{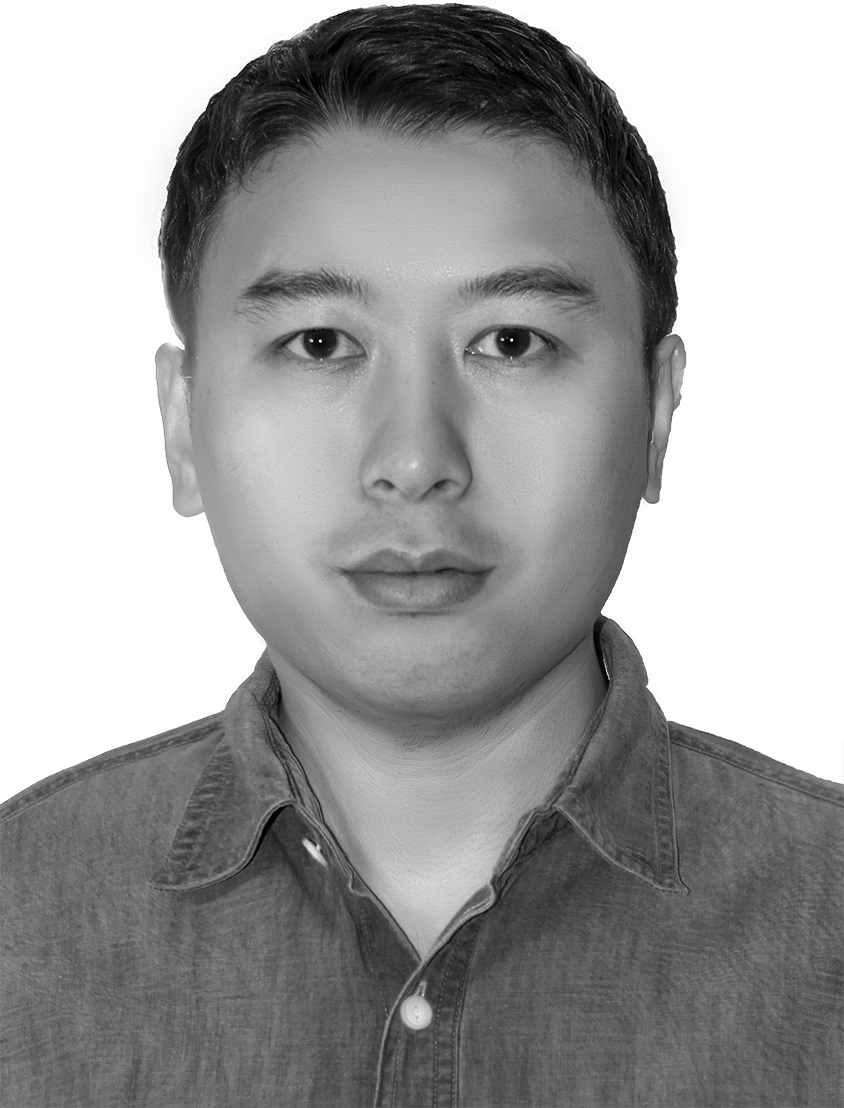}}]{HAO SHI} received the Master's degree in multimedia from Monash University, in 2014. He is currently a teacher with the School of Electronic Information, Jinzhong Vocational and Technical College of China. His main research topics include the applications of recommender systems, Ethics in Communication, natural language processing, Mathematical Statistics and the patterns of user behaviors.
\end{IEEEbiography}

\begin{IEEEbiography}[{\includegraphics[width=1in,height=1.25in,clip,keepaspectratio]{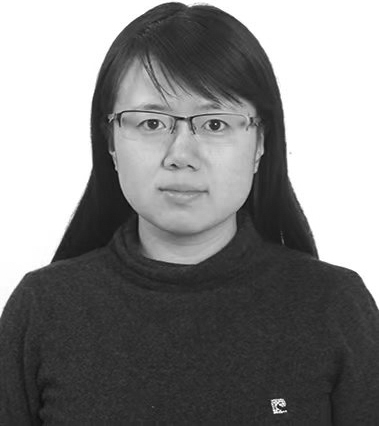}}]{ZI-JIAO WANG} received the Master’s degree in Business Administration from Shanxi University, in 2017. She is currently a civil servant with Taiyuan Tax Service, State Administration of Taxation. Her research interests include tax law and policy, digital business management, financial analysis, economic statistics , complex networks, accounting, monetary policy and recommender systems.
\end{IEEEbiography}

\begin{IEEEbiography}[{\includegraphics[width=1in,height=1.25in,clip,keepaspectratio]{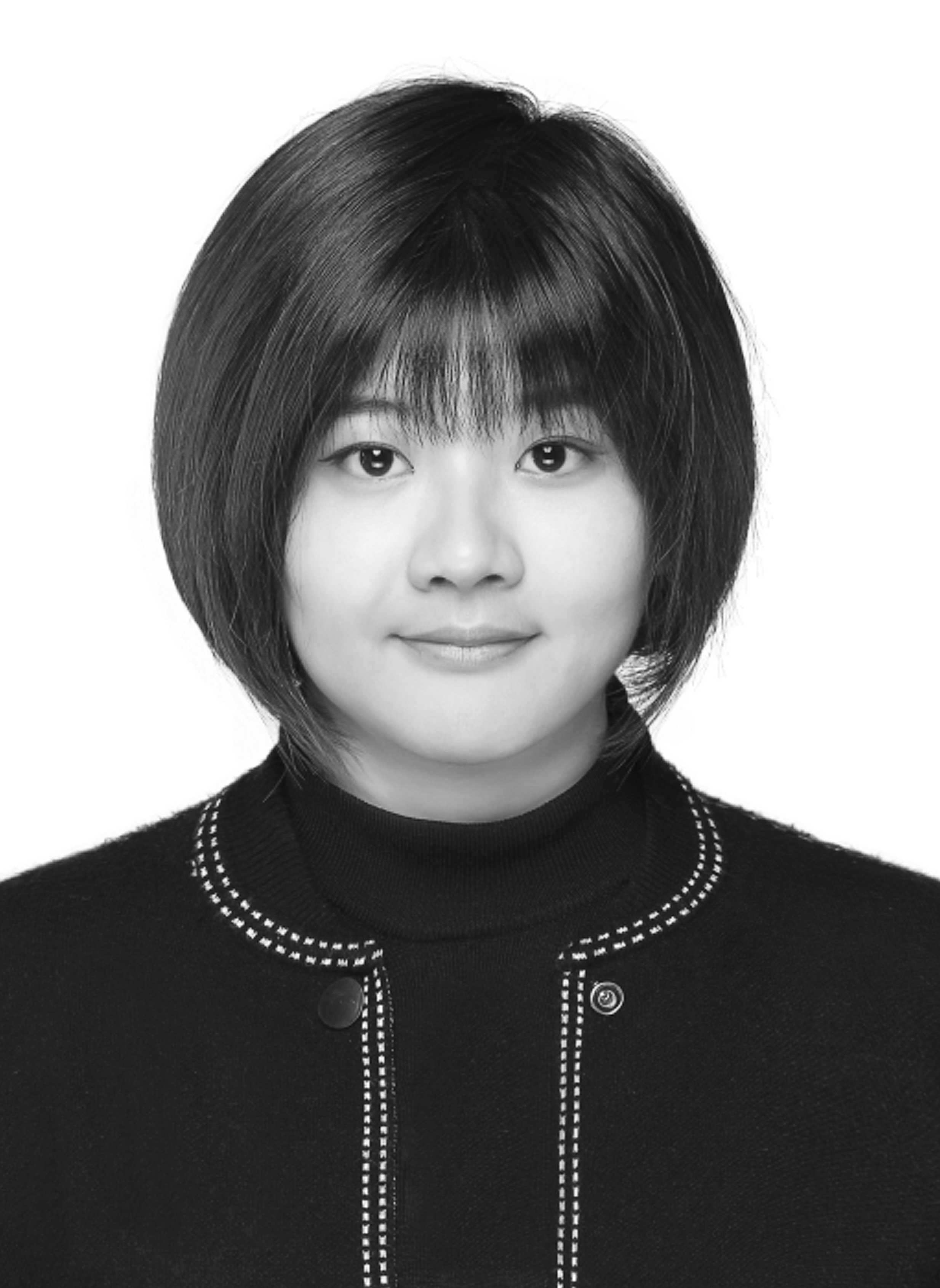}}]{LAN-RU ZHAI} received the Master’s degree in landscape garden design from North West Agriculture and Forestry University, in 2018. She currently works with the School of Electronic Information, Jinzhong Vocational and Technical College of China. Her research interests include landscape garden design, art history, digital education, data mining, computational communication and deep learning.
\end{IEEEbiography}

\Figure[t!](topskip=0pt, botskip=0pt, midskip=0pt)[width=7.16 in]{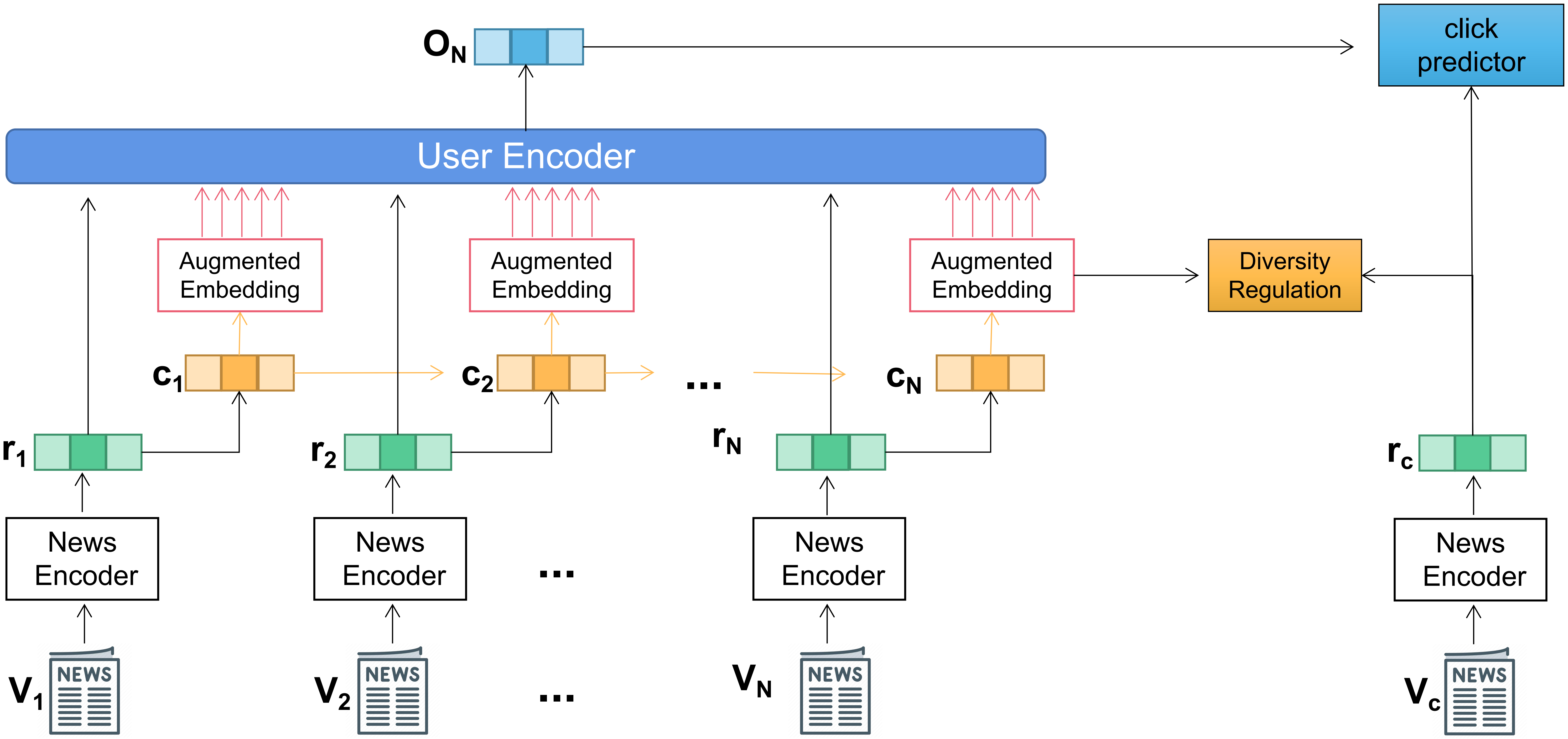}
{Model Architecture\label{fig1}}

\Figure[t!](topskip=0pt, botskip=0pt, midskip=0pt)[width=3 in]{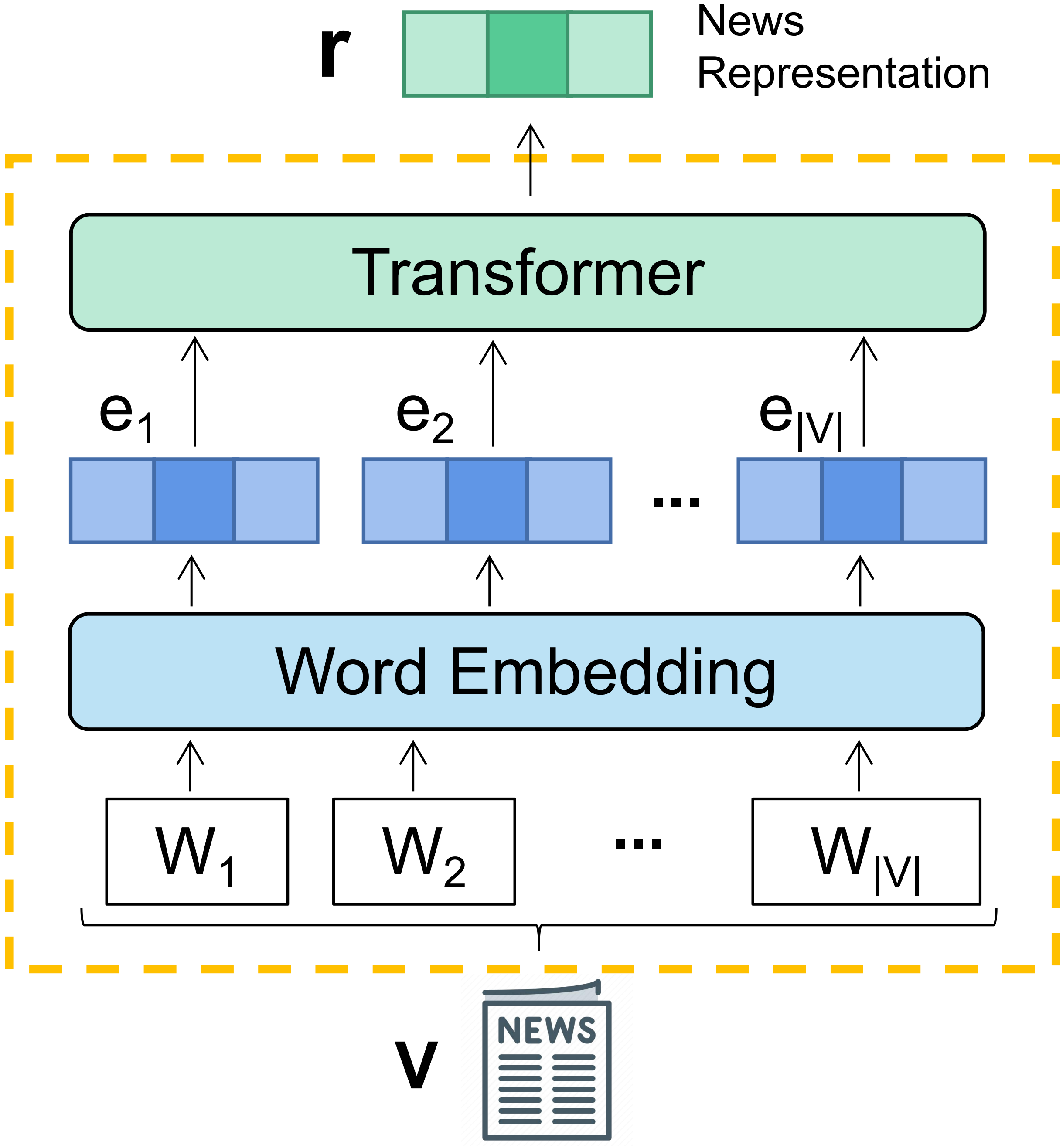}
{The framework of news encoder.\label{fig1}}

\Figure[t!](topskip=0pt, botskip=0pt, midskip=0pt)[width=7.16 in]{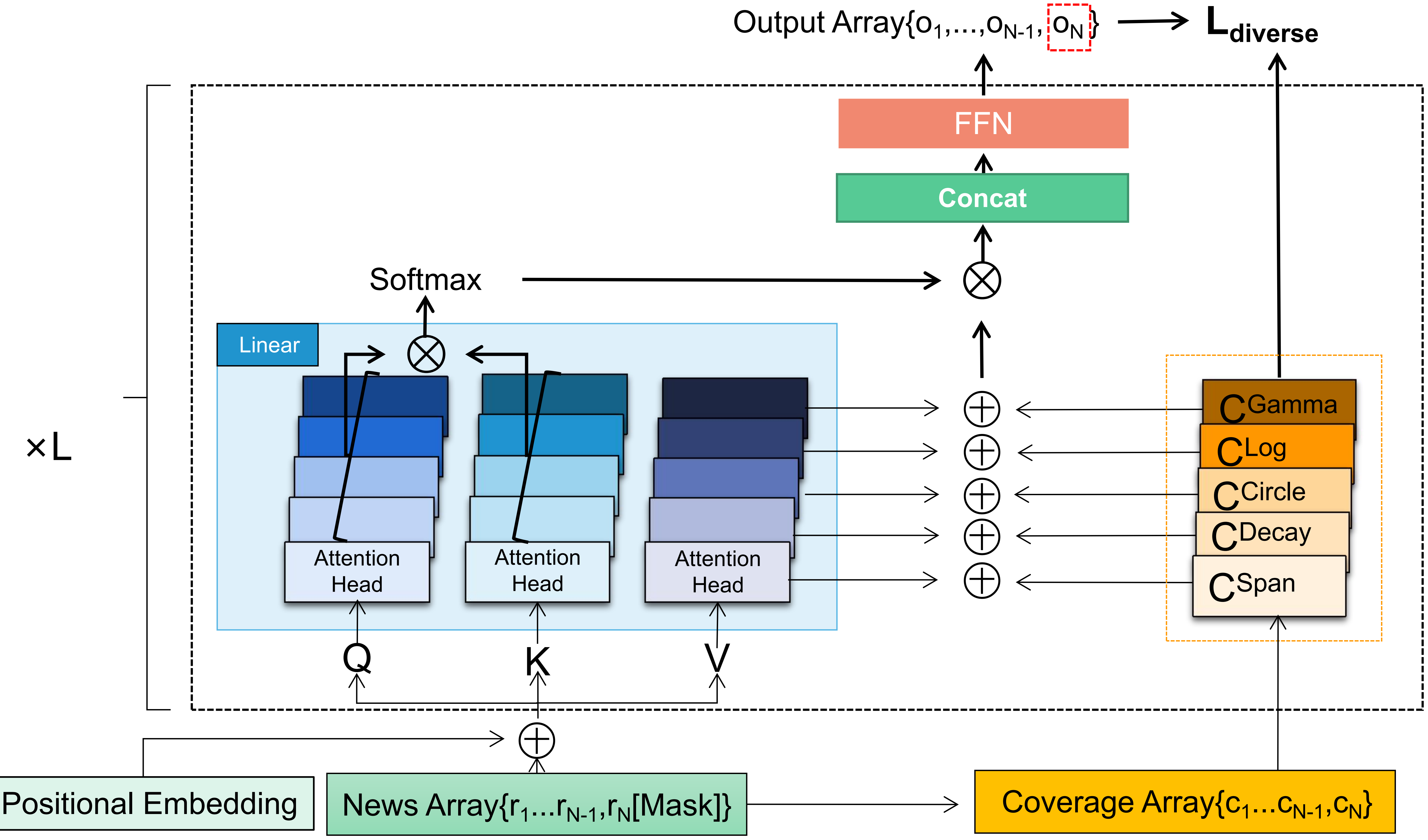}
{Brief overview of the User encoder with coverage embedded multi-headed attention structure.\label{fig1}}

\Figure[t!](topskip=0pt, botskip=0pt, midskip=0pt)[width=7.16 in]{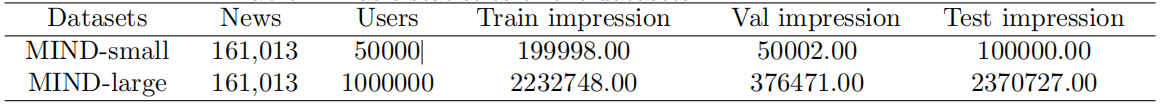}
{Basic statistics of the datasets.\label{fig1}}

\Figure[t!](topskip=0pt, botskip=0pt, midskip=0pt)[width=7.16 in]{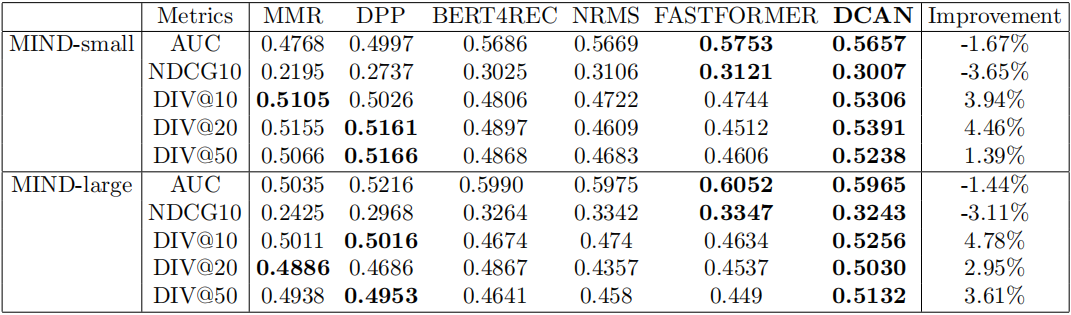}
{The accuracy and diversity performance comparison.\label{fig1}}

\Figure[t!](topskip=0pt, botskip=0pt, midskip=0pt)[width=7.16 in]{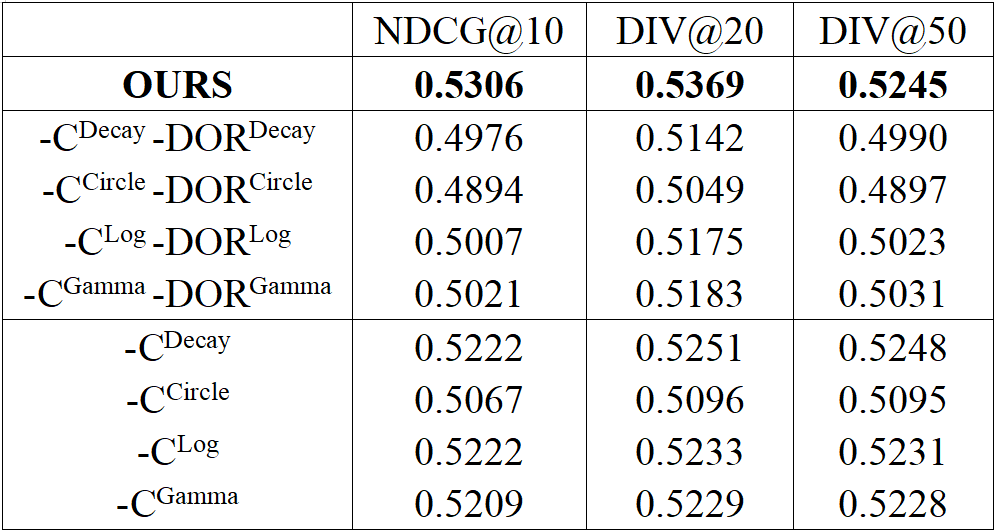}
{NDCG@10, DIV@20 and DIV@50 score of the DCAN variants.\label{fig1}}

\EOD


\begin{thebibliography}{00}

\bibitem{b1}Prawesh and B. Padmanabhan, “A complex systems perspective of news recommender systems: Guiding emergent outcomes with feedback models,” PLoS ONE, vol. 16, no. 1, p. e0245096, Jan. 2021

\bibitem{b2}Puthiya Parambath, N. Usunier, and Y. Grandvalet, “A Coverage-Based Approach to Recommendation Diversity On Similarity Graph,” \emph{in Proc. 10thACM Conf. Recomm. Syst. (RecSys)}, New York, NY, USA, 2016, pp. 15–22.

\bibitem{b3}D. P. Kingma and J. L. Ba, “Adam: A method for stochastic optimization,”in Proc. 3rd Int. Conf. Learn. Represent.(ICLR), 2015

\bibitem{b4}L. Xia, J. Xu, Y. Lan, J. Guo, W. Zeng, and X. Cheng, “Adapting Markov Decision Process for Search Result Diversification,” \emph{in Proc. 40th Int. ACM SIGIR Conf. Res. Dev. Inf. Retr.(SIGIR)}, Shinjuku Tokyo Japan, Aug. 2017, pp. 535–544.

\bibitem{b5}S. Gollapudi and A. Sharma, “An axiomatic approach for result diversification,”\emph{in Proc. 18th Int. World Wide Web Conf.(WWW)}, New York, NY, USA, 2009, pp. 381–390.

\bibitem{b6}A. Vaswani et al., “Attention is All you Need,” \emph{in Proc. Adv. neural inf. proces. syst.}, 2017, vol. 30. Accessed: May 28, 2022. 

\bibitem{b7}J. Devlin, M.-W. Chang, K. Lee, and K. Toutanova, “BERT: Pre-training of deep bidirectional transformers for language understanding,” \emph{in Proc. Conf. N. Am. Chapter Assoc. Comput. Linguistics: Hum. Lang. Technol. in Proc. Conf.(NAACL HLT) }, 2019, vol. 1,  pp.4171-4186.

\bibitem{b8}Sun et al., “BERT4Rec: Sequential Recommendation with Bidirectional Encoder Representations from Transformer,” \emph{in Proc. Int Conf Inf Knowledge Manage}, New York, NY, USA, 2019, pp. 1441–1450.

\bibitem{b9}L. Wang, C. Wang, K. Wang, and X. He, “BiUCB: A Contextual Bandit Algorithm for Cold-Start and Diversified Recommendation,” \emph{in Proc. IEEE Int. Conf. Big Knowl. (ICBK)}, 2017, pp. 248–253. 

\bibitem{b10}L. Qin, S. Chen, and X. Zhu, “Contextual Combinatorial Bandit and its Application on Diversified Online Recommendation,”\emph{in Proc. SIAM Int. Conf. Data Min.(SDM)}, Apr. 2014, pp. 461–469.

\bibitem{b11}S. Vargas, L. Baltrunas, A. Karatzoglou, and P. Castells, “Coverage, redundancy and size-awareness in genre diversity for recommender systems,” \emph{in Proc. 8th RecSys ACM Conf. Recomm. Syst.(RecSys)}, New York, NY, USA, 2014, pp. 209–216. 

\bibitem{b12}X. Zhao, L. Xia, L. Zhang, Z. Ding, D. Yin, and J. Tang, “Deep reinforcement learning for page-wise recommendations,”\emph{in Proc. 12th ACM Conf. Recomm. Syst.(RecSys)}, Vancouver British Columbia Canada, Sep. 2018, pp. 95–103.

\bibitem{b13}H. Wang, F. Zhang, X. Xie, and M. Guo, “DKN: Deep Knowledge-Aware Network for News Recommendation,”\emph{in Proc. Web Conf. World Wide Web Conf.(WWW)}, Republic and Canton of Geneva, CHE, 2018, pp. 1835–1844.

\bibitem{b14}S. Okura, Y. Tagami, S. Ono, and A. Tajima, “Embedding-based News Recommendation for Millions of Users,”\emph{In Proc. ACM SIGKDD Int. Conf. Knowl. Discov. Data Min.(KDD)}, New York, NY, USA, 2017, pp. 1933–1942. 

\bibitem{b15}C. Wu, F. Wu, T. Qi, and Y. Huang, “Empowering News Recommendation with Pre-trained Language Models,”\emph{in Proc. 44th Int. ACM SIGIR Conf. Res. Dev. Inf. Retr.}, New York, NY, USA: Association for Computing Machinery, 2021, pp. 1652–1656. Accessed: May 28, 2022. 

\bibitem{b16}J. A. Gillenwater, A. Kulesza, E. Fox, and B. Taskar, “Expectation-Maximization for Learning Determinantal Point Processes,” \emph{in Proc. Adv. neural inf. proces. syst.}, 2014, vol. 27. Accessed: May 28, 2022. 

\bibitem{b17}T. Nguyen, P.-M. Hui, F. M. Harper, L. Terveen, and J. A. Konstan, “Exploring the filter bubble: the effect of using recommender systems on content diversity,”\emph{in Proc. 23rd Int. Conf. World Wide Web.(WWW)}, New York, NY, USA, 2014, pp. 677–686.

\bibitem{b18}L. Chen, G. Zhang, and E. Zhou, “Fast Greedy MAP Inference for Determinantal Point Process to Improve Recommendation Diversity,”\emph{in Proc. Adv. neural inf. proces. syst.}, 2018, vol. 31. Accessed: May 28, 2022.

\bibitem{b19}A. See, P. J. Liu, and C. D. Manning, “Get to the point: Summarization with pointer-generator networks,” \emph{in Proc. 55th Annu. Meet. Assoc. Comput. Linguist.(ACL)}, 2017, vol. 1, pp. 1073–1083.

\bibitem{b20}J. Pennington, R. Socher, and C. Manning, “GloVe: Global Vectors for Word Representation,” \emph{in Proc. Conf. Empir. Methods Nat. Lang.(EMNLP)}, Doha, Qatar, 2014, pp. 1532–1543.

\bibitem{b21}T. Qi et al., “HieRec: Hierarchical user interest modeling for personalized news recommendation,”\emph{in Proc. 59th Annu. Meet. Assoc. Comput. Linguist. Int. Jt. Conf. Nat. Lang. Process.(ACL-IJCNLP)}, 2021, pp. 5446–5456.

\bibitem{b22}T. Silveira, M. Zhang, X. Lin, Y. Liu, and S. Ma, “How good your recommender system is? A survey on evaluations in recommendation,” \emph{Int. J. Mach. Learn. Cyber.}, vol. 10, no. 5, pp. 813–831, May 2019, doi: 10.1007/s13042-017-0762-9.

\bibitem{b23}K. Kapoor, V. Kumar, L. Terveen, J. A. Konstan, and P. Schrater, “‘I like to explore sometimes’: Adapting to Dynamic User Novelty Preferences,”\emph{in Proc. 9th ACM Conf. Recomm. Syst.(RecSys)}, New York, NY, USA, 2015, pp. 19–26.

\bibitem{b24}C.-N. Ziegler, S. M. McNee, J. A. Konstan, and G. Lausen, “Improving recommendation lists through topic diversification,” \emph{in Proc. 14th Int. Conf. World Wide Web (WWW)}, New York, NY, USA, 2005, pp. 22–32.

\bibitem{b25}K. Kapoor, K. Subbian, J. Srivastava, and P. Schrater, “Just in Time Recommendations: Modeling the Dynamics of Boredom in Activity Streams,” \emph{in Proc. 8th ACM Int. Conf. Web Search Data Min.(WSDM)}, New York, NY, USA, 2015, pp. 233–242.

\bibitem{b26}J. L. Ba, J. R. Kiros, and G. E. Hinton, “Layer Normalization.” arXiv, Jul. 21, 2016. doi: 10.48550/arXiv.1607.06450.

\bibitem{b27}S. Li, Y. Zhou, D. Zhang, Y. Zhang, and X. Lan, “Learning to Diversify Recommendations Based on Matrix Factorization,” \emph{in Proc. 15th IEEE Int. Conf. Dependable, Auton. Secur. Comput., IEEE Int. Conf. Pervasive Intell. Comput., IEEE Int. Conf. Big Data Intell. Comput. IEEE Cyber Sci. Technol. Congr., DASC-PICom-DataCom-CyberSciTec}, 2017, pp. 68–74.

\bibitem{b28}P. Cheng, S. Wang, J. Ma, J. Sun, and H. Xiong, “Learning to Recommend Accurate and Diverse Items,”\emph{in Proc. Int. World Wide Web Conf.(WWW)}, Republic and Canton of Geneva, CHE, 2017, pp. 183–192.

\bibitem{b29}M. Gartrell, U. Paquet, and N. Koenigstein, “Low-rank factorization of determinantal point processes,”\emph{in Proc. 31st AAAI Conf. Artif. Intell.(AAAI)}, USA, 2017, pp. 1912–1918.

\bibitem{b30}R. M. Cho, E. Park, and S. Yoo, “MEANTIME: Mixture of Attention Mechanisms with Multi-temporal Embeddings for Sequential Recommendation,” \emph{in Proc. 14th ACM Conf. Recomm. Syst.(RecSys)}, Sep. 2020, pp. 515–520.

\bibitem{b31}F. Wu et al., “MIND: A Large-scale Dataset for News Recommendation,” \emph{in Proc. Annu. Meet. Assoc. Comput Linguist.}, Online, 2020, pp. 3597–3606.

\bibitem{b32}Z. Tu, Z. Lu, L. Yang, X. Liu, and H. Li, “Modeling coverage for neural machine translation,”\emph{in Proc. 54th Annu. Meet. Assoc. Comput. Linguist.(ACL)}, vol. 1, pp. 76–85.

\bibitem{b33}X. He, L. Liao, H. Zhang, L. Nie, X. Hu, and T.-S. Chua, “Neural Collaborative Filtering,” \emph{in Proc. 26th Int. World Wide Web Conf.(WWW)}, Perth Australia, Apr. 2017, pp. 173–182.

\bibitem{b34}C. Wu, F. Wu, S. Ge, T. Qi, Y. Huang, and X. Xie, “Neural News Recommendation with Multi-Head Self-Attention,” \emph{in Proc. 9th Conf. Empir. Methods Nat. Lang. Process. Int. Jt. Conf. Nat. Lang. Process.(EMNLP-IJCNLP)}, Hong Kong, China, 2019, pp. 6389–6394.

\bibitem{b35}J. Hendrickx, A. Smets, and P. Ballon, “News Recommender Systems and News Diversity, Two of a Kind? A Case Study from a Small Media Market,” \emph{Journalism and Media}, vol. 2, no. 3, Art. no. 3, Sep. 2021

\bibitem{b36}E. Gabrilovich, S. Dumais, and E. Horvitz, “Newsjunkie: providing personalized newsfeeds via analysis of information novelty,” \emph{in Proc. 13th Int. World Wide Web Conf. Proc. (WWW)}, New York, NY, USA, 2004, pp. 482–490.

\bibitem{b37}M. Reuver, A. Fokkens, and S. Verberne, “No NLP Task Should be an Island: Multi-disciplinarity for Diversity in News Recommender Systems,” \emph{in Proc. 16th EACL Hackashop News Media Content Anal. Autom. Rep. Gener., Hackashop Conf. Eur. Chapter Assoc. Comput. Linguist.(EACL)}, Online, 2021, pp. 45–55. Accessed: May 28, 2022.

\bibitem{b38}C. Wu, F. Wu, M. An, J. Huang, Y. Huang, and X. Xie, “NPA: Neural News Recommendation with Personalized Attention,”\emph{in Proc. ACM SIGKDD Int. Conf. Knowl. Discov. Data Min.(KDD)}, New York, NY, USA, 2019, pp. 2576–2584.

\bibitem{b39}P. Adamopoulos and A. Tuzhilin, “On over-specialization and concentration bias of recommendations: probabilistic neighborhood selection in collaborative filtering systems,” in Proceedings of the 8th ACM Conference on Recommender systems, New York, NY, USA, 2014, pp. 153–160. 

\bibitem{b40}J. Tang and K. Wang, “Personalized Top-N Sequential Recommendation via Convolutional Sequence Embedding,” in Proceedings of the Eleventh ACM International Conference on Web Search and Data Mining, New York, NY, USA, 2018, pp. 565–573.

\bibitem{b41}T. Qi, F. Wu, C. Wu, and Y. Huang, “PP-Rec: News recommendation with personalized user interest and time-aware news popularity,”\emph{in Proc. 59th Annu. Meet. Assoc. Comput. Linguist. 11th Int. Jt. Conf. Nat. Lang. Process.(ACL-IJCNLP)}, 2021, p. p 5457-5467.

\bibitem{b42}L. Qin and X. Zhu, “Promoting diversity in recommendation by entropy regularizer,” \emph{in Proc. Int. Joint Conf. Artif. Intell.(IJCAI)}, Beijing, China, 2013, pp. 2698–2704.

\bibitem{b43}L. Li, D. Wang, T. Li, D. Knox, and B. Padmanabhan, “SCENE: a scalable two-stage personalized news recommendation system,” \emph{in Proc. 34th Int. ACM SIGIR Conf. Res. Dev. Inf. Retr.(SIGIR)}, New York, NY, USA, 2011, pp. 125–134. 

\bibitem{b44}W. -C. Kang and J. McAuley, “Self-Attentive Sequential Recommendation,” \emph{in Proc. IEEE Int. Conf. Data Min. (ICDM)}, 2018, pp. 197–206.

\bibitem{b45}C. Wu, F. Wu, T. Qi, and Y. Huang, “SentiRec: Sentiment Diversity-aware Neural News Recommendation,” \emph{in prof. 1st conf. Asia-Pacific Chapter Assoc. Comput. Linguist 10th Int. Jt. Conf. Nat. Lang. Process.}, Suzhou, China, 2020, pp. 44–53. 

\bibitem{b46}J. Rao, A. Jia, Y. Feng, and D. Zhao, “Taxonomy Based Personalized News Recommendation: Novelty and Diversity,”\emph{in Proc. Lect. Notes Comput. Sci.(WISE)}, Berlin, Heidelberg, 2013, pp. 209–218.

\bibitem{b47}R. Warlop, J. Mary, and M. Gartrell, “Tensorized Determinantal Point Processes for Recommendation,”\emph{in Proc. ACM SIGKDD Int. Conf. Knowl. Discov. Data Min.(KDD)}, New York, NY, USA, 2019, pp. 1605–1615.

\bibitem{b48}P. Liu, K. Shivaram, A. Culotta, M. A. Shapiro, and M. Bilgic, “The Interaction between Political Typology and Filter Bubbles in News Recommendation Algorithms,” \emph{in Proc. Web Conf. World Wide Web Conf.(WWW)}, New York, NY, USA, 2021, pp. 3791–3801.

\bibitem{b49}J. Li, Y. Wang, and J. McAuley, “Time Interval Aware Self-Attention for Sequential Recommendation,” \emph{in Proc. 13th Int. Conf. Web Search Data Min.(WSDM)}, New York, NY, USA, 2020, pp. 322–330. Accessed: May 28, 2022.

\bibitem{b50}Z. Dai, Z. Yang, Y. Yang, J. Carbonell, Q. V. Le, and R. Salakhutdinov, “Transformer-XL: Attentive language models beyond a fixed-length context,”\emph{in Proc. 57th Annu. Meet. Assoc. Comput. Linguist.}, 2020, pp 2978-2988.

\bibitem{b51}C. Wu, F. Wu, T. Qi, and Y. Huang, “Two Birds with One Stone: Unified Model Learning for Both Recall and Ranking in News Recommendation.” \emph{arXiv}, Mar. 23, 2022. Accessed: May 28, 2022. 

\bibitem{b52}Q. Zhang et al., “UNBERT: User-News Matching BERT for News Recommendation,” \emph{in Proc. Int. Joint Conf. Artif. Intell.(IJCAI)}, Montreal, Canada, Aug. 2021, pp. 3356–3362.

\bibitem{b53}C. Wu, F. Wu, T. Qi, and Y. Huang, “User Modeling with Click Preference and Reading Satisfaction for News Recommendation,”\emph{in Proc. Int. Joint Conf. Artif. Intell.(IJCAI)}, Yokohama, Japan, Jul. 2020, pp. 3023–3029.

\bibitem{b54}Z. Yang, Z. Dai, Y. Yang, J. Carbonell, R. R. Salakhutdinov, and Q. V. Le, “XLNet: Generalized Autoregressive Pretraining for Language Understanding,”\emph{in Proc. Adv. neural inf. proces. syst.}, 2019, vol. 32. Accessed: May 28, 2022.

\bibitem{b55}C. Wu, F. Wu, Y. Huang, and X. Xie, “Personalized News Recommendation: Methods and Challenges,” \emph{ACM Trans. Inf. Syst.}, 2022

\bibitem{b56}G. Adomavicius and Y. Kwon, “Improving Aggregate Recommendation Diversity Using Ranking-Based Techniques,”\emph{IEEE Trans Knowl Data Eng}, vol. 24, no. 5, pp. 896–911, 2012

\bibitem{b57}G. Zheng et al., “DRN: A deep reinforcement learning framework for news recommendation,” \emph{in Proc. Web Conf. World Wide Web Conf.(WWW)}, 2018, pp. 167-176.

\bibitem{b58}C. Wu, F. Wu, T. Qi, Y. Huang, and X. Xie, “Fastformer: Additive attention can be all you need,” \emph{arXiv}, 2021

\end{thebibliography}
\end{document}